# Distributed Clustering for User Devices Under Unmanned Aerial Vehicle Coverage Area during Disaster Recovery

Abdu Saif [1], kaharudin bin dimyati [1], Kamarul Ariffin Bin Noordin [1], Nor Shahida Mohd.Shah[2], S. H. Alsamhi[3], Qazwan Abdullah[2] and Nabil Farah[4]

[1]Faculty of Engineering, Department of Electrical Engineering, University of Malaya,50603, Kuala Lumpur, Malaysia
[2]Faculty of Engineering Technology, Universiti Tun Hussein Onn Malaysia, Pagoh, Muar, Johor, Malaysia.
[3]Software Research Institute, Athlone Institute of Technology, Athlone, Ireland & Faculty of Engineering,
IBB University, Ibb, Yemen
[4]Faculty of Electrical and Electronic Engineering Technology, Universiti Teknikal Malaysia Melaka.

*Abstract*—An Unmanned Aerial Vehicle (UAV) is a promising technology for providing wireless coverage to ground user devices. For all the infrastructure communication networks destroyed in disasters, UAVs- battery life is challenging during service delivery in a post-disaster scenario. Therefore, selecting cluster heads among user devices plays a vital role in detecting UAV signals and processing data for improving UAV energy efficacy and reliable Connectivity. This paper focuses on the performance evaluation of the clustering approach performance in detecting wireless coverage services with improving energy efficiency. The evaluation performance is a realistic simulation for the ground-to-air channel Line of Sight (LoS). The results show that the cluster head can effectively link the UAVs and cluster members at minimal energy expenditure. The UAV's altitudes and path loss exponent affected user devices for detecting wireless coverage. Moreover, the bit error rate in the cluster heads is considered for reliable Connectivity in post-disaster. Clustering stabilizes the clusters linking the uncovered nodes to the UAV, and its effectiveness in doing so resulted in its ubiquity in emergency communication systems.

## I. INTRODUCTION

Natural disasters include earthquakes, hurricanes, tornadoes, and severe snowstorms, resulting in devastating telecommunication infrastructures. Natural disasters affect the economy, damage infrastructure, the environment, humanitarian crises, and death. The effects solutions include delivering communication from space technology (i.e., satellite, high altitude platform, low altitude platform) and advanced digital technologies such as Artificial Intelligence (AI), Blockchain technology, robotics, etc. Unmanned Arial Vehicles (UAV) represent the efficient solution to deliver communication services, providing food, gathering data, guiding Search, and Rescue team (SAR)[1], [2]. UAV technology currently plays an essential role in combating COVID-19 pandemics via delivering food and goods, monitoring and spray disinfection, and detecting infected cases from long distances [3]. In such a case, the user devices cannot establish the access link with the terrestrial network wireless network and obtain the coverage services. UAVs can be used as a mobile base station to provide wireless coverage services in disaster-stricken areas. This allows emergency communication services to continue uninterrupted despite the damaged infrastructures [2]. UAVs in disaster-stricken areas increase the overall wireless coverage while minimizing the channel access delays during natural disasters. Space technologies are an efficient way of disaster recovery after all terrestrial communications are destroyed [4]. The UAVs can function at low and high altitudes and provide a Line of Sight (LoS) communication link to user devices[5]. Its transmission and distance coverages establish reliable connections at minimal energy expenditure[6],[7]. The authors of [8] addressed how UAV is used to gather data from IoT devices to improve energy efficiency. Simultaneously, the authors of [9] discussed the efficient techniques and technology for securing UAVs due to their sensitive applications in the real world.

Furthermore, improving the Quality of Services (QoS) depends on LoS and received signal strength and bandwidth, throughput, delay, etc [10]. UAVs integrated with Device to Device (D2D) communications keep communication lines open and running during the chaos of natural disasters such as earthquakes, floods, and tsunamis. Gathering data during a disaster is essential to guide research and rescue teams to perform their complex tasks efficiently [11].

An internet portfolio of technologies uses IoT to improve the integration between device-oriented sensor networks and data-oriented applications. UAVs and D2D communications suffer from power consumption caused by standardization and handling disaster-resilient communication [12]. Previous works on UAVs and D2D communications focused on IoT platforms that enable novel solutions for items that have a significant qualitative effect on people's lives, such as smart cities, smart grids, smart homes, and connected vehicles [13]. These studies highlighted the need for IoTs due to their rapidly gaining ground in future wireless communications, such as 5G [14],[15].

Clustering provides efficient and stable routes for data dissemination. It links user devices via direct communication to improve network performance for communicative data sharing [16],[17]. In a natural disaster, ground user devices cannot obtain wireless coverage services due to damaged infrastructures [18]. Cluster Heads (CHs) are anodes that detect wireless coverage services transmitted by the UAV and forward it to the Cluster Members (CMs) via downlinks [12],[19]. This allows CHs to minimize the UAV's overload and increase communicative efficiency in a post-disaster scenario. CHs establish communication links with CMs based on the D2D communication pair within the cluster's shot range area. Clustering is convenient and reliable, energy-efficient, and useful during emergencies. Communication is critical during natural disasters as it allows ground user devices to make contact with the outside world.

This work focuses on UAV for providing coverage services integrated the clustering approach for users based on D2D communication when the telecommunication infrastructure is damaged due to natural disasters. The cluster head detects wireless coverage services from UAVs and establishes communication links with other users' devices within a disaster zone. The network's energy lifetime increases while the UAV's and D2D's load will be minimized due to communication clustering. This approach increases the coverage of the D2D due to its streamlining of the Connectivity and efficiency of post-disaster communication.

## II. SYSTEM MODEL

The scenario of post-disaster communication is shown in Fig.1 The UAVs deploy to provide wireless coverage services to Ground User Devices (GUDs) in the case of infrastructure collapse. The UAV can achieve mobility in the disaster zone and carry the radio coverage services to provide the CHs wireless coverage services to improve services coverage and sustainable Connectivity. Besides, the UAV smartness, intelligence for controlling and communicating with the CHs.

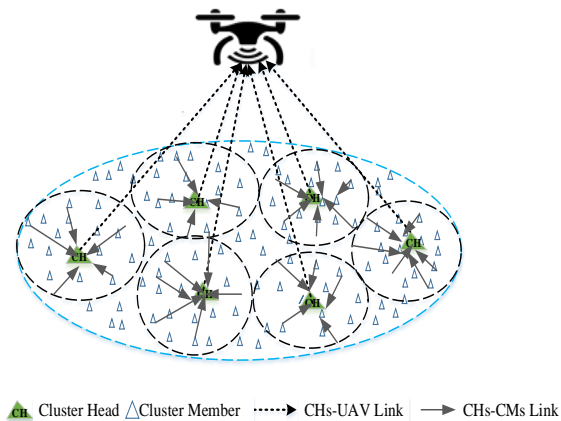

Fig. 1. CHs and user devices under the UAV coverage area

The distributed GUDs are based on the Poisson Point Clustering (PPC) and D2D communications architecture. The GUDs contain residual energy that exceeds a threshold and can obtain communication links with the UAV acting as a cluster head. $CH_i$ is responsible for obtaining wireless coverage from the UAV, where $i = 1,2.........N$. The GUDs, which contain residual energy below the threshold, connect within short-range with CHs to get wireless coverage services. D2D communication establishes between the CH and CMs within the cluster range. CH communicates with the UAV and CMs in full-duplex mode to send/receive data packets. The proposed model aims to verify the QoS signals' reliability and availability on GUD receivers during natural disasters by minimizing the communication link and increasing the system capacity. The UAVs fly autonomously and provide wireless coverage services during disaster recovery and are integrated with clustering nodes, and D2D communication can support fifth-generation (5G) wireless communications. Besides, UAVs are location-dynamic and respond to an emergency independently via quick reconfiguration for effective communications and disaster recovery.

Fig.2 shows the CHs sensing wireless coverage from the UAV and its forwarding to the CMs via the D2D communication link. The Ground to Air (G2A) channel communication link was evaluated using an uplink communication. The CHs senses the wireless coverage signals from the UAV then forwards them to the CMs based on the clustering within the D2D communication links. In this context, two scenarios are analyzed for transmitting wireless communication during disasters; the CHs-to-UAV link is represented by scenario (I), while the communication link of CH-to-CMs is represented in scenario (II). In these scenarios, the UAV is configured to detect the CHs to transmit data packets and maximize the CHs nodes throughput for effective Connectivity. The CHs act as central distribution nodes for the CMs; the wireless coverage received in full-duplex mode improves the communication link between the clustering network via D2D communication. The CHs are expected to provide a more efficient and stable route solution to the system post-disasters, minimize the communication loads for UAV, and do so at minimal energy expenditure.

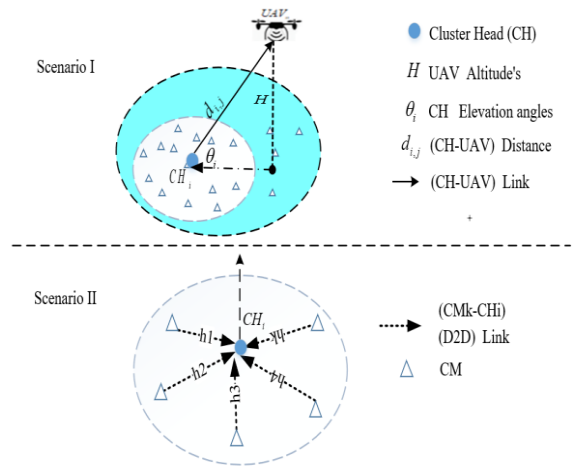

Fig. 2. Clustering and Communication

The LoS communication link was established between the cluster head and UAV in the uplink. CHs detects the wireless coverage signals and selects which UAV connection that the wireless coverage services link to the CHs will forward the coverage services to the CMs in a D2D communication scenario. A G2Achannel was established between the CHs and UAV, while the ground to ground ( G2G) channel was established between CH and CMs.

The optimal CHs can communicate with the UAV to obtain wireless coverage services. The UAV's altitude is designed to maximize the coverage probability and provide reliable Connectivity based on the position of the CHs during disasters. The UAVs adjust their attitudes in line with the distance for optimal CHs to guarantee a communication link. It was assumed that the CHs is located in $(x_i, y_i)$, CMs is located in $(x_k, y_k)$, and UAV is located in $(x_j, y_j, z_j)$ in the 3D- space. The distance between the CHs and UAV are calculated as follows:

$$d_{i,j} = \sqrt{(x_i - x_j)^2 + (y_i - y_j)^2 + (z_i - z_j)^2} \quad (1)$$

The distance between the CH to CMs are:

$$d_{i,k} = \sqrt{(x_i - x_k)^2 + (y_i - y_k)^2} \quad (2)$$

The probability of an LoS's uplink between the CHs and UAV are denoted by $P_{Los}(\varphi)$ As follows [12]:

$$P_{Los}(i,j) = \frac{1}{1 + ae^{-b(\theta_i - a)}} \quad (3)$$

where $\theta_i$ is the elevation angle of the CHs, and $a$, and $b$ are parameters affecting the S-curve parameters that vary based on the environment such as urban, suburban, dense urban, and high-rise urban. From Eq (3), the link is more likely to be an LoS communication link with a larger elevation angle.

The elevation angle for the CHs to sensing the wireless coverage single for the UAV depends on the UAV altitudes H and distance d from CHs to UAV.

$$\theta_i = \frac{180}{\pi} arcsin(\frac{H}{d}) \quad (4)$$

From (3), the LoS probability increases as the elevation angle increases. The elevation angle also increases as per the distance and altitudes of the UAV. In this context, it was assumed that the CHs could detect wireless coverage services from UAV based on the LoS probability at levels exceeding the threshold (ε close to 1):

$$PLos \geq \varepsilon \Rightarrow \theta_i \geq P_{Los}^{-1}(\varepsilon) \quad (5)$$

The evolution angle should lead to:

$$d_{i,j} \leq \frac{h_j}{sin(P_{Los}^{-1}(\varepsilon))} \quad (6)$$

Eq. (6) shows that the range of coverage distance between the CHs and UAV should meet the connection conditions. The UAV received power nodes, j, from CH node, i or j, is given by [13] (in dB):

$$p_r^{i,j} = p_t^i - 10\alpha \, log\left(\frac{4\pi f_c d_{i,j}}{c}\right) - \eta \quad (7)$$

Where $p_t^i$ is the power of transmission of CHs in dB, $f_c$ is the carrier frequency, $\alpha = 2$ is the path loss exponent for LoS propagation, $\eta$ is an excessive path loss added to the free space propagation loss, and $c$ is the speed of light. The minimum transmit power of CHs required to reach the UAV for reliable connectivity, and bit error rate requirement is denoted as:

$$p_t^{i,j} = [Q^{-1}(\delta)]^2 \frac{R_b \sigma^2}{2} 10^{\eta/10} \left(\frac{4\pi f_c d_{i,j}}{c}\right)^2 \quad (8)$$

Where $Q^{-1}(.)$ is the inverse Q-function, $\sigma^2$ is the noise power spectral density, and $R_b$ is denoted as the transmission bit rate. When deriving (8) using (7), the bit error rate expression of the received power at the UAV was found based on the Quadrature Phase Shift Keying (QPSK):

$$p_e = Q \sqrt{\frac{2p_r^{i,j}}{R_b \sigma^2}} \quad (9)$$

The objective is to detect efficient, reliable Connectivity in a disaster zone with the lowest possible energy expenditure. The UAV was configured to respond to the CHs, and a minimum signal-Noise Ratio (SNR) is needed to decode the gathered signals successfully.

TABLE I. SIMULATION PARAMETERS [13]

| Parameters | Description | Value |
|---|---|---|
| a, b | Urban environment parameter | 9.6 ,0.16 |
| $f_c$ | Carrier frequency | 3.5 GHz |
| $p_t^i$ | CH transmission power | 13 dB |
| $\delta$ | Bit error rate requirement | $10^{-8}$ |
| $\varepsilon$ | $P_{LoS}$ Requirement | 0.95 |
| $\sigma^2$ | Noise power spectral density | -170 dBm/Hz |
| $R_b$ | Transmission data rate | 200 Kbps |
| B | Transmission bandwidth per device | 200 kHz |
| $\eta$ | Additional path loss to free space | 5 dB |

III. RESULTS AND DISCUSSION

The simulation results (parameters in Table 1) are analyzed the CHs and UAV's G2A connectivity for efficient disaster communications. The distributed CHs within the disaster zone detect the UAV wireless coverage and then forward it to the CMs. We consider the distance between CHs-to UAV's (500 m) and receives the former's signals via uplink communication. The D2D communication between CMs and CHs is made possible via the cluster's residual energy. The probability LoS, UAV received signals, and the bit error rate is evaluated based on the CHs-UAV's distance, CHs elevation angles, and path loss exponents.

The results have been presented to verify the reliability and the availability of QoS signals on GUD receivers during natural disasters. In this context, the GUDS obtain the wireless coverage services in the short-range communication through the CHs.

That will help minimize the power consumption and interference between the GUDs. The Clustering approach will also reduce the signals traffic load of UAV and increase the system capacity.

Fig 3. shows that the LoS probability decreases when the CHs-UAV distance increases simultaneously due to the changes in coverage level across UAV altitudes. It can also be seen that LoS's maximum probability is achieved at a distance of 120 m.

The elevation angle is 90º between the CHs and UAV in the vertical coverage services. When the distance exceeds 120 m, the $P_{LoS}$ decreases drastically due to the decreased received SNR with increasing distances. The altitudes increased to 150 m and 200 m, the $P_{LoS}$ decreased from its maximum value write number please at a distance 200 m and 250 m due to the increased coverage range versus altitude from 100 m to 200 m and the decreasing signal strengths. The reason is that UAVs altitude can gain height and hover over a particular region, and mainly operate within the LoS range of the receiver.

The UAVs act to increase their gain, fly over a region, and operate optimally within the receiver's LoS range from the CHs. The CHs detect UAVs' wireless coverage signals and forward them to the wireless coverage of the CMs. The LoS link's probability represents the ground user devices.

Conversely, Fig 4. shows that the LoS probability's performance increases when the CHs elevation angle increases across multiple UAV altitudes. The minimum probability of LoS is achieved at an elevation angle of 0º at the height of 100 m and increases to 1 when the CHs elevation angle increased to 70º at altitudes of 150 m and 200 m. Rician fading occurs when one path typically receives signals at low altitudes associated with LoS uncongested with traffic or strong reflection signals.

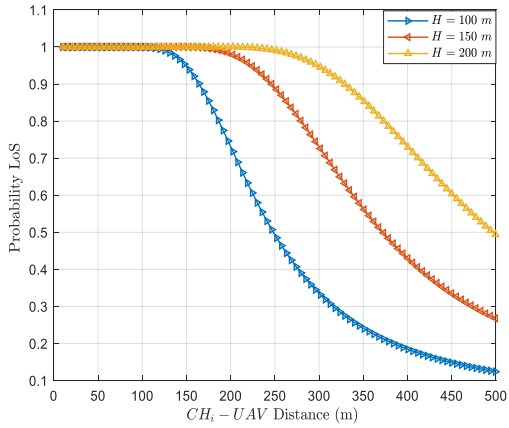

Fig. 3. Probability LoS Versus the CHs- UAV distance with different altitudes.

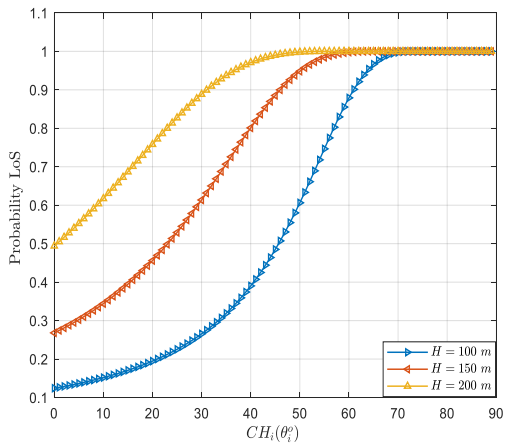

Fig. 4. Probability LoS Versus the CHs Elevation Angle with different altitudes.

Fig. 5 shows the CHs elevation angles affected by the CHs-UAV distance in the targeted altitudes. The CHs elevations angle decreased while the CHs-UAVs length increased for different altitudes due to the reduced LoS wireless coverage services. For H=100 m, the range of the elevation angle versus distance was 45º to 12º, while at H=150 m, the range of effective CHs elevation angle was 45º - 16º. However, at H=200 m, the CHs elevation angle decreased or increased from 45º to 20º. Furthermore, the CHs elevation angle decreased with increasing distance and UAV altitudes due to decreased LoS probability and increased massive scale path loss. Finally, lower altitudes and small-scale fading were found in the A2G channel instead of a G2G link, while having a longer link length deteriorates the received SNR.

Fig. 6 shows the received signal's performance at the nodes versus the elevation angle of the CHs with multiple path loss exponents. Data transfer from CHs can be delivered to the UAVs via the uplink channel. Therefore, it can be surmised that the UAVs' received signal from the $i^{th}$ ground CHs was affected by the elevation angle and path loss exponent.

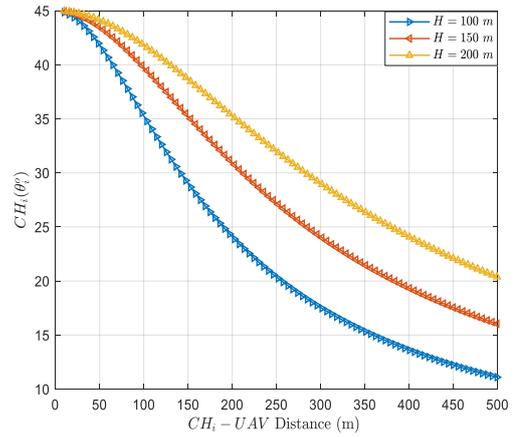

Fig. 5. CHs elevation ranges versus CHs -UAV distance with different altitudes.

At α = 2, the received signal increased from -220 dB to -150 dB in the same elevation angle of CHs. At α =2.5, the received signal increased from -250 dB to -180 dB, and at α =3, the received signal increased from -310 dB to -220 dB. The α affects the SNR at the UAV via the massive-scale path loss linked with the transmission distance between the UAV and CHs. The path loss exponent negatively influences the UAV-received SNR large-scale path loss due to distance.

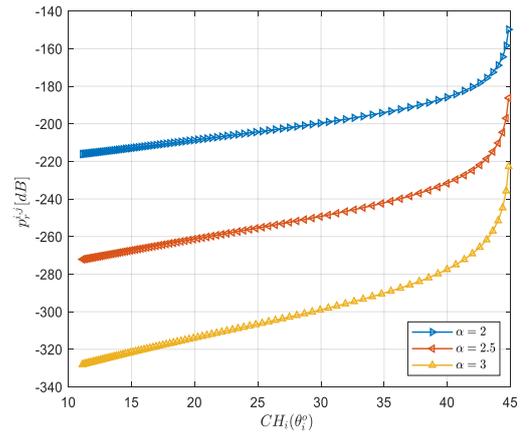

Fig. 6. UAV received signals versus CHs elevation Angle with different $\alpha$.

Fig 7. shows the performance of the bit error rate versus the received signal at the UAV with multiple α. It can be seen that the bit error rate achieved at the massive sale path loss $\alpha$ =3 due to decreased SNR at the receiver nodes. However, the low bit error rate is seen at $\alpha$ =2 due to the received signal

only from the LoS at the destination nodes and small-scale path loss that effectively receive signals. Therefore, the bit error rate also increased at $\alpha$ =2.5 for the same level CHs elevation angles.

The bit error rate increased versus the CH selection at different α due to the massive scale path loss and small path loss. The clustering performance, D2D communication network, and UAV network elucidates during post-disaster scenarios. The network capacity measures the amount of traffic that every network scenario will handle at a minimum bit error rate post-disaster.

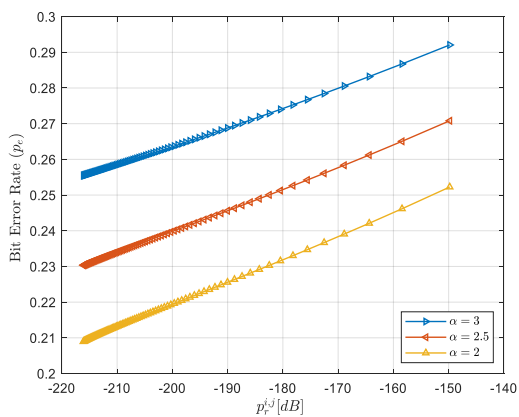

Fig.7. Uplink bit error rate versus UAV received single.

IV. CONCLUSIONS

This paper focuses on the clustering approach for users under the UAV coverage area during disaster recovery. CHs are elected to detect UAVs' wireless coverage for providing emergency coverage services in disaster-stricken areas. UAV power consumption decreases when establishing a communications link visible to the ground users. Then, CHs shows a communication link with CMs and D2D communication with the UAV decentralized control. The performance of the probability of LoS and received signals is evaluated at multiple UAV altitudes to ensure its effectiveness in post-disaster communications. The collaboration of multi UAV for delivering communication services during disaster recovery can be considered for further future contribution integrated distributed clustering for the user in the disaster area.

ACKNOWLEDGEMENT

This work is supported by the University of Malaya, Faculty of Electrical Engineering, and funded by the DARE project (Grand ID: IF035A-2017 & IF035-2017 ) and Universiti Tun Hussein Onn Malaysia under the TIER1 Grant Vot H243.